\documentclass[multphys,vecphys]{svmult}

\usepackage{makeidx}         % allows index generation
\usepackage{graphicx}        % standard LaTeX graphics tool
\usepackage{multicol}        % used for the two-column index
\usepackage[bottom]{footmisc}% places footnotes at page bottom

\makeindex             % used for the subject index
\newcommand{\coone}{$\rm ^{12}CO(1-0)$}
\newcommand{\cotwo}{$\rm ^{12}CO(2-1)$}

\newcommand{\ea}{et al.}

\newcommand{\myr}{\>{\rm Myr}}

\newcommand{\pc}{\>{\rm pc}}
\newcommand{\kpc}{\>{\rm kpc}}

\newcommand{\msun}{\>{\rm M_{\odot}}}
\newcommand{\lsun}{\>{\rm L_{\odot}}}

\newcommand{\as}{^{\prime\prime}}
\newcommand{\am}{^{\prime}}
\newcommand{\bdm}{\begin{displaymath}}
\newcommand{\edm}{\end{displaymath}}
\newcommand{\beq}{\begin{equation}}
\newcommand{\eeq}{\end{equation}}
\newcommand{\bit}{\begin{itemize}}
\newcommand{\eit}{\end{itemize}}
\newcommand{\ben}{\begin{enumerate}}
\newcommand{\een}{\end{enumerate}}
\newcommand{\bfi}{\begin{figure}[htb]}
\newcommand{\bpfi}{\begin{figure}[p]}

\begin{document}

\title*{Nuclear Star Clusters across the Hubble Sequence}

\author{Torsten B\"oker}

\institute{European Space Agency, Dept. RSSD, Keplerlaan 1, 
2200 AG Noordwijk, Netherlands
\texttt{tboeker@rssd.esa.int} }

\maketitle

\begin{abstract}
Over the last decade, HST imaging studies have revealed that the centers of most galaxies are
occupied by compact, barely resolved sources. Based on their structural properties, position in the
fundamental plane, and spectra, these sources clearly have a stellar origin. They are therefore called
``nuclear star clusters'' (NCs) or ``stellar nuclei''. NCs are found in galaxies of all Hubble types,
suggesting that their formation is intricately linked to galaxy evolution. In this contribution, 
I briefly review the results from recent studies of NCs, touch on some ideas for their
formation, and mention some open issues related to the possible connection between NCs and
supermassive black holes.
\end{abstract}

\vspace*{-5mm}

\section{Introduction} \label{sec:intro}
\vspace*{-3mm}
The nuclei of galaxies are bound to provide ``special'' physical conditions because they 
are located at the bottom of the potential well of their host galaxies. This unique
location manifests itself in various distinctive phenomena such as active galactic nuclei
(AGN), central starbursts, or extreme stellar densities. The evolution 
of galactic nuclei is closely linked to that of their host galaxies, as inferred from a 
number of global-to-nucleus scaling relations discovered in the last decade.

Recently, observational and theoretical interest has been refocused
onto the compact and massive star clusters found in the nuclei of galaxies of all Hubble 
types. Historically, the nuclei of dE,N galaxies have been best studied, but it has become 
clear from recent HST studies that similar objects exist also in 
normal spirals and ellipticals. At face value, these ``nuclear star 
clusters\footnote{A word on terminology: historically, the compact light excess found 
in the centers of ``nucleated'' early-type galaxies is often referred to as a 
``stellar nucleus''. In most studies of disk-dominated galaxies, however, the 
term ``nuclear star cluster'' is used. Because we now know that in terms of 
size and luminosity, both types of objects are indistinguishable, and likely 
differ only their in evolutionary stage, it seems reasonable to
adopt a single term for them. Since ``star cluster''
seems more descriptive of the nature of these objects than ``nucleus''
(after all, every galaxy has a nucleus, i.e. a center), we will refer to 
them as ``nuclear star clusters'' or NCs for all galaxies types.}''
are an intriguing environment for the formation of massive black holes because 
of their extreme stellar density. NCs may also constitute the progenitors of at least some 
halo globular clusters via ``NC capture'' following the tidal disruption 
of a satellite galaxy. Finally, their formation process is influenced by (and 
important for) the central potential, which in turn governs the
secular evolution of their host galaxies.
In what follows, I briefly summarize what has been learned about NCs over the last few years.

\begin{figure}[t]
\centering
\includegraphics[width=7cm]{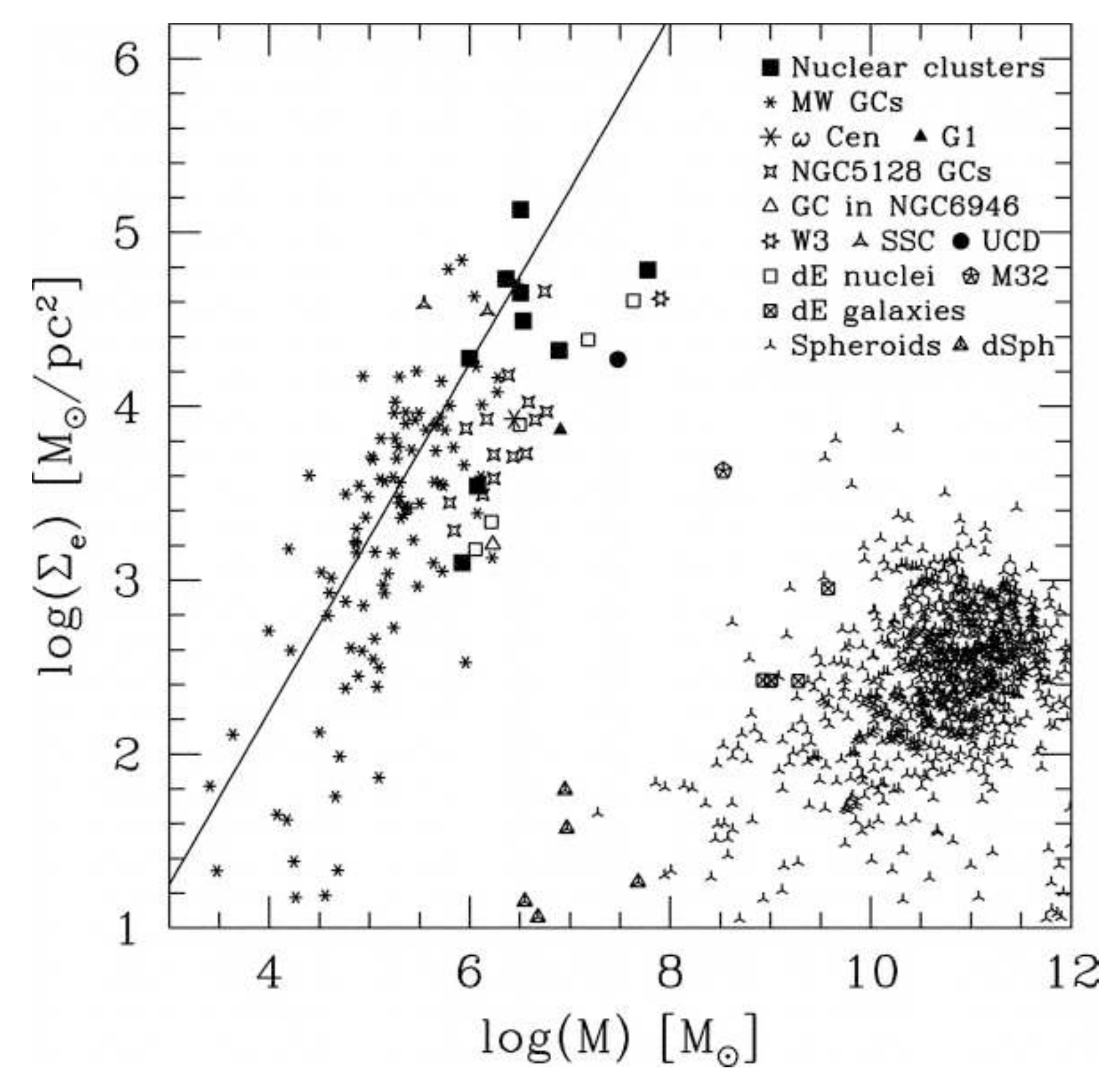}
\caption{Mean projected mass density of various stellar systems inside their 
effective radius $r_e$, plotted against their total mass.
This is similar to a face-on view of the fundamental plane. 
NCs occupy the high end a region populated by other types of massive stellar clusters,
and are well separated from elliptical galaxies and spiral bulges. The solid line 
represents a constant cluster size, i.e. $r_e=3\pc$ (from \cite{wal05}).}
\label{fig:fp}
\end{figure}

\section{Properties of Nuclear Star Clusters} \label{sec:properties}
\vspace*{-3mm}
Extragalactic star clusters are compact sources, and in general, their study 
requires space-based resolution. It doesn't come as a surprise, therefore, that
the HST has been instrumental for recent progress in the understanding of
NCs. Over the last decade, a number of HST studies - both via imaging and
spectroscopic observations - have contributed to the following picture of NCs:

\vspace*{2mm}
\noindent {\bf 1)} NCs are common: the fraction of galaxies with an unambiguous NC detection
is 75\% in late-type (Scd-Sm) spirals \cite{boe02}, 50\% in earlier-type (Sa-Sc)
spirals \cite{car97}, and 70\% in spheroidal (E \& S0) galaxies \cite{cot06}. All these
numbers are likely lower limits, although for different reasons. In the latest-type
disks, it is sometimes not trivial to locate the galaxy center unambiguously so that
no particular source can be identified with it. In contrast, many early-type galaxies
have very steep surface brightness profiles (SBPs) that make it difficult to detect
even luminous clusters against this bright background.

\vspace*{2mm}
\noindent {\bf 2)} NCs are much more luminous than ``normal'' globular clusters (GCs). With typical
absolute I-band magnitudes between -14 and -10 \cite{boe02,cot06}, they are roughly 40 times more 
luminous than the average MW globular cluster \cite{har96}.

\vspace*{2mm}
\noindent {\bf 3)} However, NCs are as compact as MW GCs. Their half-light radius typically is $\rm 2-5\,pc$, 
independent of galaxy type \cite{boe04,geh02,cot06}. 

\vspace*{2mm}
\noindent {\bf 4)} Despite their compactness, NCs are very massive: their typical dynamical mass is 
$10^6 - 10^7\msun$ \cite{wal05} which is at the extreme high end of the GC mass function. 

\vspace*{2mm}
\noindent {\bf 5)} Their mass density clearly separates NCs from compact galaxy bulges. This
is demonstrated in Figure~\ref{fig:fp} which compares the mass and mass density of
NCs to that of other spheroidal stellar systems. The clear gap between bulges/ellipticals
on the one hand, and NCs on the other hand makes a direct evolutionary connection
between the two classes of objects unlikely.

\vspace*{2mm}
\noindent {\bf 6)} The star formation history of NCs is complex, as evidenced by the fact
that most NCs have stellar populations comprised of multiple generations of stars
\cite{wal06,ros06}. The youngest generation is nearly always younger than $100\myr$ which
is strong evidence that NCs experience frequent and repetitive star formation episodes
\cite{wal06}.

\begin{figure}[t]
\centering
\includegraphics[width=7cm]{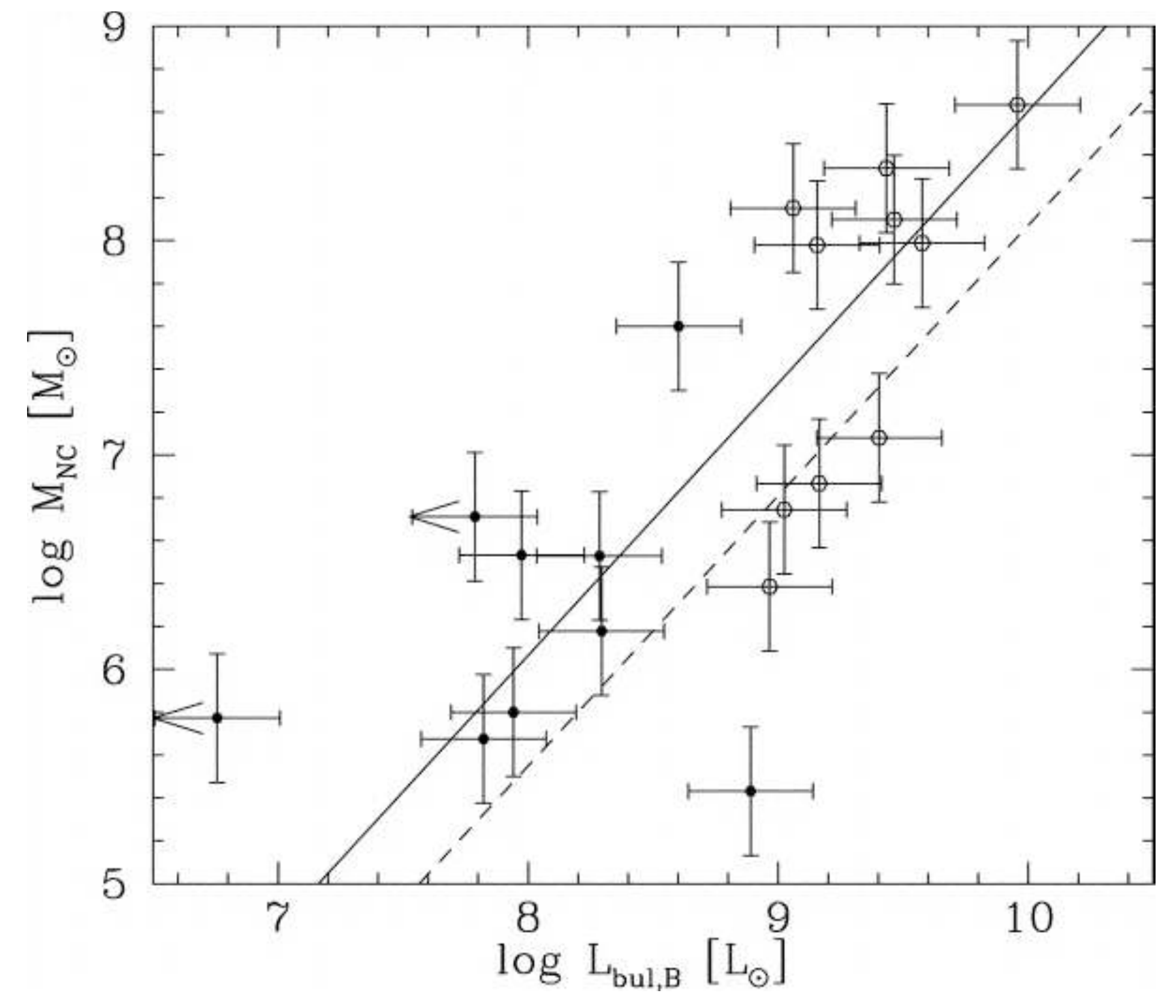}
\caption{Relation between NC mass in spiral galaxies and bulge luminosity $\rm log(L_B)[\lsun ]$ 
of the host galaxy (from \cite{ros06}). Open symbols denote early-type spirals, and filled 
symbols denote late-type spirals. There is a strong correlation in the sense that galaxies 
with more luminous bulges have more massive NCs. The solid line indicates the best 
linear fit to the data, while the dashed line indicates the relation between the SMBH mass
and bulge luminosity for the sample of \cite{mar03}.}
\label{fig:rossa}
\end{figure}

\vspace*{2mm}
\noindent {\bf 7)} Due to three recent and independent studies of NCs in different
galaxy types \cite{ros06,weh06,cot06}, it
has become clear that NCs obey similar scaling relationships with host galaxy 
properties as do supermassive black holes (SMBHs). As an example, Figure~\ref{fig:rossa}
shows the NC mass as a function of bulge luminosity. While the implications 
of this result are not yet clear (see \S~\ref{sec:para}), these studies have renewed interest 
in NCs because of the potentially important role that NCs play in the evolution of their
host galaxies.

\section{How (and when) do Nuclear Clusters Form?} \label{sec:formation}
\begin{figure}
\centering
\includegraphics[width=12cm]{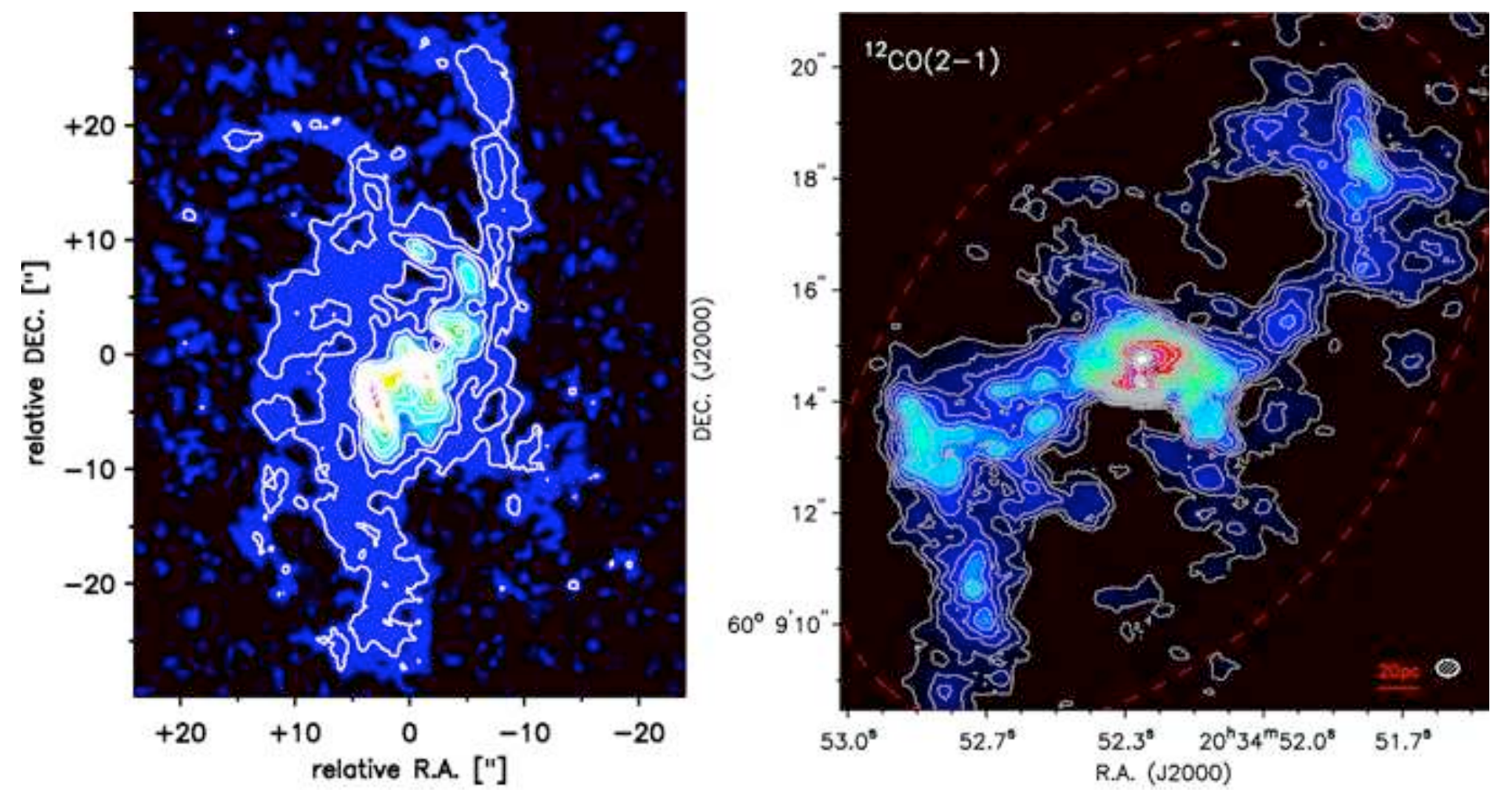}
\caption{Left: intensity map of the \coone\ line in the central 
$1\am$ ($1.6\kpc$) in the nearby spiral galaxy NGC\,6946 \cite{sch06}.
Right: higher resolution ($0.4\as$) map of the \cotwo\ emission
in the central $300\pc$. Note the S-shaped spiral that appears to ``funnel''
molecular gas towards the central $10\pc$ where about $1.6\cdot 10^7\msun$ of
molecular gas have accumulated \cite{sch07}.}
\label{fig:n6946} 
\end{figure}

\vspace*{-3mm}
The processes that funnel gas onto NCs in the local universe have recently been
studied in some detail, enabled by significant improvements to the sensitivity
and spatial resolution of mm-interferometers. Figure~\ref{fig:n6946} shows the 
molecular gas distribution in the nearby spiral NGC\,6946.
Both the morphology and the kinematics of the gas can be well explained by the
effects of a small-scale stellar bar. The S-shaped flow pattern onto the
nucleus and the large ($1.6\cdot 10^7\msun$) gas concentration in the inner 
$\approx 10\pc$ lend credibility to the ``repetitive burst'' scenario for NC growth.

Less clear, however, are the reasons for why gas accumulates in the nucleus of a shallow
disk galaxy {\it in the absence} of a prominent central mass concentration, i.e.
how the ``seed clusters'' form initially. A few studies have attempted to provide 
an explanation for this puzzle. For example, \cite{mil04} suggests the magneto-rotational 
instability in a differentially rotating gas disk as a viable means to transport gas 
towards the nucleus and to support (semi)continuous star formation there.

More recently, \cite{ems07} have pointed out that the tidal field becomes compressive
in shallow density profiles, causing gas to collapse onto the nucleus of a disk galaxy. 
If correct, then NC formation is indeed expected to be a natural consequence of galaxy 
formation, which would go a long way towards explaining at least some of the observed
scaling relations between NCs and their host galaxies.

The question of when a particular NC (i.e. its ``seed'' cluster) has formed is 
equivalent to asking how old its oldest stars are. This question is extremely 
difficult to answer in all galaxy types, albeit for different reasons. In late-type 
spirals, for example, the NC nearly always contains a young stellar population
which dominates the spectrum and thus makes the detection of an underlying
older population challenging, not to mention its accurate age determination.

Early-type, spheroidal galaxies, on the other hand, lack the large gas 
reservoirs of spirals, and thus should experience less frequent nuclear starbursts.
One therefore would expect their NCs to contain fewer and older stellar populations.
However, early-type galaxies have much steeper surface brightness profiles, and therefore
a low contrast between NC and galaxy body. This makes spectroscopic studies of 
NCs in E's and S0's exceedingly difficult.
The few published studies have focused on the NCs of dE,N galaxies, and have
shown that even these can have rather young (few hundred Myrs) stars, as demonstrated 
e.g. by \cite{but05} in the case of NGC\,205.
Generally speaking, however, most NCs in dE,N galaxies have integrated colors that -  
while different from those of their host galaxies - are generally consistent with evolved 
stellar populations at least 1\,Gyr old \cite{sti01}. Considering that there may be even
older stellar populations ``hiding'', this age should only be considered a 
lower limit for the oldest stellar population in dE,N nuclei. 

In fact, it is not implausible that the ``seed clusters'' for present-day NCs were 
in place very early in the universe. The average star formation rate over the last
100\,Myr in NCs of late-type spirals is $\rm 2\cdot 10^{-3}\msun /yr$ \cite{wal06}.
Assuming this SFR was constant for the past 10\,Gyr, one would expect a stellar mass
of $\approx 2\cdot 10^7\msun$ which is within a factor of 4 from the typical NC mass of 
$5\cdot 10^6\msun$ \cite{wal05}. Turning the argument around, if NCs indeed build up
their entire present-day mass via a series of repetive starbursts, then they must have 
been in place at least 3\,Gyr ago, unless their SFRs were significantly higher in the 
past than over the last 100\,Myr. Given that observations todate are somewhat biased 
towards more luminous NCs which likely have a time-averaged SFR higher than the
``typical'' NC, this estimate might even be too low.

\section{A new Paradigm?} \label{sec:para}
\vspace*{-3mm}
It has recently been proposed by \cite{fer06} that NCs extend the well-known scaling 
relation between the mass of a galaxy and that of its central super-massive black 
hole (SMBH) to lower masses. This has triggered speculation about
a common formation mechanism of NCs and SMBHs, being governed mostly by
the mass of the host galaxy. The idea put forward is that NCs and SMBHs are
two incarnations of a ``central massive object'' (CMO) which forms in
every galaxy. In galaxies above a certain mass threshold ($\approx 10^{10}\msun$),
galaxies form predominantly SMBHs while lower mass galaxies form NCs.

While tantalizing, this apparent connection opens more questions than it answers.
For example, we know that some galaxies contain {\it both} an NC and a SMBH. A 
well-known example is the ``mini-Seyfert'' NGC\,4395 \cite{fil03}, but others have been found
recently \cite{sat07,shi07}. Why then do some NCs contain SMBHs, but not all? 
Why do some galaxies apparently contain {\it neither} NC nor SMBH? Is an NC possibly 
a pre-requisite for the formation of a SMBH? Is the formation of a BH (not necessarily 
a super-massive one) a logical consequence of the high stellar densities present
in NCs? Progress along these lines will require a better understanding of the 
formation of ``pure'' disk galaxies in the early universe, 
as well as improved models for the evolution of extremely dense stellar
systems.

\printindex
\end{document}